\documentclass[aps,prb,twocolumn,superscriptaddress,longbibliography]{revtex4-2}
 
\usepackage[T1]{fontenc}
\usepackage{graphicx}
\usepackage{amsmath}
\usepackage{amssymb}
\usepackage{float}
\usepackage{chemformula}
\usepackage[version=4]{mhchem}
\usepackage{multirow}

\begin{document}
 
\title{Continuous $\mathcal{PT}$-Symmetry Breaking as a Design Variable for Giant Altermagnetic Spin Splitting}
 
\author{Kichan Chun}
\affiliation{Department of Physics, Sejong University, Seoul 05006, Korea}
 
\author{Gunn Kim}
\email{gunnkim@sejong.ac.kr}
\affiliation{Department of Physics, Sejong University, Seoul 05006, Korea}
 
\date{\today}
 
\begin{abstract}
Magnetic point-group analysis classifies altermagnets but returns only a binary symmetry verdict, leaving spin-splitting energy (SSE) inaccessible without spin-polarized density functional theory (DFT). This binary ceiling is not fundamental. Sublattice symmetry breaking is promoted here to a continuous, DFT-free scalar---the Motif Symmetry-Breaking Index (MSBI)---that quantifies $\mathcal{PT}$-symmetry breaking between antiparallel magnetic motifs directly from crystal coordinates. SHAP analysis of an XGBoost surrogate trained on 3,851 DFT-labeled binary structures identifies three dominant descriptors: MSBI (symmetry-breaking axis), motif packing fraction MPF (superexchange axis), and the $p/d$ electron ratio (covalency axis), each mapping onto a directly tunable experimental handle. A controlled VO--CrSb comparison within the same P$6_3$/mmc host lattice demonstrates that composition alone boosts SSE sevenfold. Bayesian optimization over this three-axis space, followed by independent DFT validation, recovers $\alpha$-NiS (SSE $= 0.823$\,eV) as cross-validation against an independent symmetry-based prediction and identifies three previously unrecognized high-SSE candidates---square-planar FeS (1.297\,eV), octahedral CoS (1.103\,eV), and FeAs (1.089\,eV)---all matching or exceeding CrSb. Square-planar Fe--S is proposed as a transferable coordination motif for giant altermagnetic spin splitting, advancing altermagnet design from symmetry classification to continuous quantitative optimization.
\end{abstract}


\maketitle

\section{Introduction}

A central challenge in the rational design of altermagnetic materials is that the theory capable of classifying them and the theory capable of quantifying them operate on incompatible mathematical footings. Altermagnets form a third class of collinear magnets in which the two antiparallel spin sublattices are connected not by spatial inversion or translation, as in a conventional antiferromagnet, but by a rotational point-group symmetry of the crystal\cite{vsmejkal2020crystal,vsmejkal2022beyond,jungwirth2025altermagnetism,mazin2022altermagnetism}. This local distinction has a dramatic electronic consequence: the material carries zero net magnetization, eliminating the stray-field constraints that limit ferromagnets\cite{vzutic2004spintronics,hirohata2020review}, yet generates momentum-dependent spin splittings that can exceed 1\,eV with characteristic $d$-wave, $g$-wave, or higher-order symmetries\cite{vsmejkal2022emerging,hayami2019momentum,bhowal2024ferroically}. Recent experimental confirmations in \ce{CrSb}\cite{reimers2024direct,ding2024large,krempasky2024altermagnetic,osumi2024observation} and \ce{MnTe}\cite{lee2024broken} via spin-resolved ARPES, together with observations of the anomalous Hall effect\cite{feng2022anomalous,peng2025scaling,bai2025nonlinear} and spin-current generation in zero-magnetization compounds\cite{kluczyk2024coexistence,li2025fully}, have established altermagnetism as a concrete materials-physics opportunity for dense, field-free spintronic architectures\cite{dong2025field,kapri2025spin,song2025altermagnets,bai2024altermagnetism}. The rigorous theoretical criterion for altermagnetism is now well-established: magnetic point-group analysis determines whether a crystal's sublattice relations permit momentum-dependent spin splitting\cite{cheong2025altermagnetism,radaelli2024tensorial}, and Yuan and Zunger have recast these formal conditions into a real-space rule based on the interconvertibility of spin-structure motif pairs\cite{yuan2023degeneracy}. What both frameworks return, however, is binary. They cannot say whether a symmetry-allowed candidate will exhibit an SSE of 10\,meV or 1\,eV---the distinction that governs functional utility. The quantity that matters for device performance is continuous; the tools that predict it are discrete. Bridging this gap currently requires full spin-polarized density functional theory (DFT) on every candidate, creating a circular dependency in which evaluation is nearly as costly as discovery.

\begin{figure*}[ht]
    \centering
    \includegraphics[width=\textwidth]{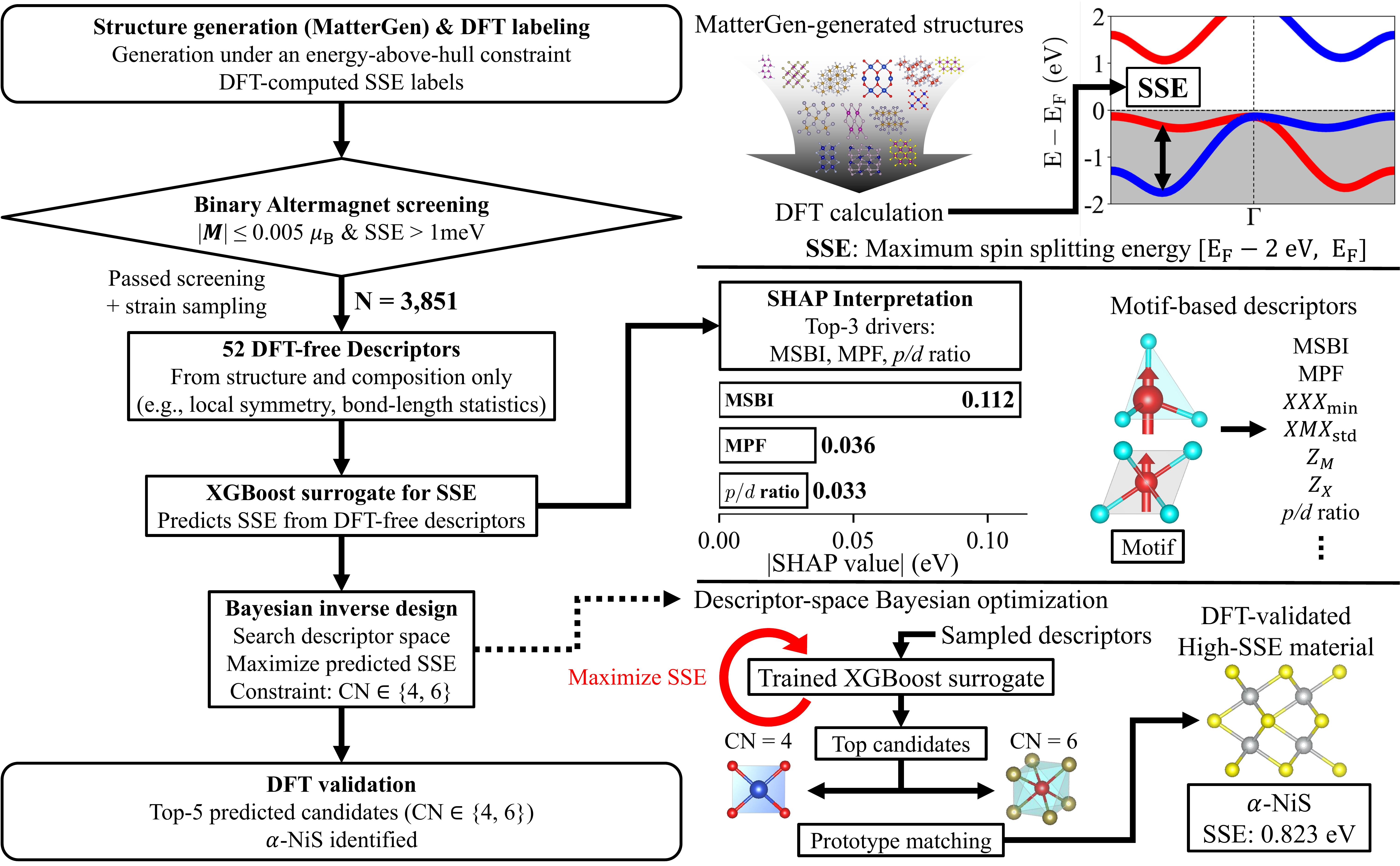}
    \caption{Overview of the inverse-design workflow for high-SSE altermagnetic materials. (Left) Candidate structures are generated with MatterGen under an energy-above-hull constraint and labeled with DFT-computed SSE, defined as the maximum spin splitting within $[E_F - 2~\text{eV},\, E_F]$. After binary-altermagnet screening ($|\boldsymbol{M}| \leq 0.005~\mu_\text{B}$, $\mathrm{SSE} > 1~\mathrm{meV}$) and strain sampling ($N = 3{,}851$), 52 motif-level descriptors are constructed entirely from crystal structure and composition, without any electronic-structure input. An XGBoost surrogate trained on these descriptors predicts SSE, and SHAP analysis identifies MSBI, MPF, and the $p/d$ electron ratio as the three leading contributors. (Right) Descriptor-space Bayesian optimization then maximizes the predicted SSE; top candidates are stratified by coordination number (CN $\in \{4,6\}$) and mapped onto periodic structures through prototype matching for independent DFT validation.}
    \label{fig:workflow}
\end{figure*}

The structural consequences of this discrete--continuous mismatch are visible in the present state of the field. Experimentally confirmed altermagnets remain confined to a small number of structural families---NiAs-type \ce{CrSb}\cite{reimers2024direct,li2025large,rai2025direction,urata2024high} and \ce{MnTe}\cite{amin2024nanoscale,liu2024chiral}, rutile \ce{MnF2}\cite{morano2025absence}, and metallic \ce{KV2Se2O}\cite{jiang2025metallic}---and recent high-throughput screenings have largely remained at the level of enumerating known structure types rather than generating new ones\cite{guo2023spin,sodequist2024two,gao2025ai}. In this data-scarce regime, architectures requiring $10^4$--$10^5$ labeled examples to generalize across independent chemistries cannot be deployed reliably; interpretable low-dimensional models are therefore not a compromise against accuracy but a principled choice. The legitimate target of machine learning here is to identify the minimal set of continuous, physically interpretable axes on which the symmetry-breaking physics lives, and to make those axes directly accessible to experimental control.

Here we show that altermagnet design admits such a continuous reformulation, organized by three complementary descriptors. The Motif Symmetry-Breaking Index (MSBI), a permutation-invariant scalar, measures in the fixed crystal frame how far the two magnetic sublattice motifs deviate from the two spatial relations---identity and inversion---whose presence would protect $\mathcal{PT}$-induced spin degeneracy. MSBI is constructed as the product of two dimensionless deviations: the motif direct deviation $D_d$, a bond-length-normalized RMSD between the ligand-vector sets of the two sublattices optimized via the Hungarian algorithm\cite{kuhn1955hungarian}, and the motif inversion deviation $D_I$, computed after point-inverting one ligand set. The multiplicative form ensures that MSBI vanishes whenever either $\mathcal{PT}$-protecting relation is approximately satisfied and is amplified only when both are simultaneously broken. The result is a continuous, differentiable scalar with a sharp empirical threshold---MSBI $> 0.5$---above which SSE systematically exceeds 0.4\,eV. This is the central claim of this work: symmetry breaking, the very quantity that gates altermagnetism, can itself be treated as a smooth design variable. Two additional axes complete the design space. The motif packing fraction (MPF) sets the superexchange coupling scale through the hopping-integral dependence $J \sim t^2/U$\cite{anderson1959new,coey2010magnetism}; the ligand-to-metal $p/d$ electron ratio encodes covalency within the Zaanen--Sawatzky--Allen framework\cite{zaanen1985band}. All three depend only on crystal structure and elemental identity, keeping the surrogate fully DFT-free at inference while exposing experimentally tunable handles---lattice distortion, applied pressure, and chemical substitution. SHAP analysis\cite{lundberg2017unified} of an XGBoost surrogate\cite{chen2016xgboost} trained on 3,851 MatterGen-generated\cite{zeni2025generative} and DFT-labeled binary structures confirms that these three descriptors dominate SSE prediction, and a controlled comparison between VO and CrSb---which share the same P$6_3$/mmc host lattice---isolates the covalency axis: composition alone drives a sevenfold SSE enhancement, from 0.165 to 1.194\,eV.

To test whether these axes are actionable, we couple the surrogate to Bayesian optimization and subject the top candidates to independent spin-polarized DFT validation (\textbf{Figure~1}). Among six magnetically stable candidates, NiAs-type $\alpha$-\ce{NiS} (SSE $= 0.823$\,eV, 8.5\% surrogate error) provides quantitative cross-validation: its altermagnetic character has been identified independently through symmetry-based bonding arguments\cite{mandal2025deterministic}, and its NiAs structure type is accessible by established synthesis routes\cite{gore2024growth,sun2014phase,tan2019solvothermal,liu2014controlled}. The primary predictive output consists of three candidates with no prior altermagnetic identification: square-planar FeS (1.297\,eV), octahedral CoS (1.103\,eV), and octahedral FeAs (1.089\,eV), all matching or exceeding the SSE of \ce{CrSb} ($\sim$1.2\,eV)\cite{reimers2024direct,li2025large}. Square-planar FeS is most productively read as a \emph{transferable motif hypothesis}: the square-planar Fe--S coordination environment itself supports giant altermagnetic splitting, wherever it can be stabilized.


\section{Methods}
\label{sec:methods}

The computational framework presented here builds on the author's master's thesis\cite{Chun2025thesis}; the present work refines the descriptor definitions, introduces the controlled MSBI construction without Kabsch alignment, and adds the Bayesian optimization pipeline for inverse design.

\subsection{DFT Calculations}
\label{subsec:dft_details}

Spin-polarized DFT calculations were performed with VASP\cite{kresse1993ab,kresse1996efficiency,kresse1996efficient} using PAW pseudopotentials\cite{blochl1994projector,kresse1999ultrasoft}, the PBE functional\cite{perdew1996generalized}, and a 500~eV plane-wave cutoff. The Brillouin zone was sampled on $\Gamma$-centered meshes with spacing $<\!0.13$~\AA$^{-1}$, and electronic convergence was set to $10^{-6}$~eV. Structures were relaxed in two steps: full cell and ionic relaxation ($\text{ISIF}=3$) followed by ionic refinement ($\text{ISIF}=2$), with forces converged below $10^{-2}$~eV/\AA. On-site Coulomb interactions were treated within the Dudarev GGA+$U$ formalism\cite{dudarev1998electron}. The element-dependent $U_{\mathrm{eff}}$ values listed in Table~\ref{tab:Uvalues} are consistent with previously proposed parameter sets for transition-metal oxides and intermetallics\cite{cococcioni2005linear,jain2011formation,kirklin2015open}. All calculations used collinear antiparallel initial moments and neglected spin--orbit coupling, consistent with the exchange-driven origin of altermagnetic splitting\cite{vsmejkal2022beyond}.

\begin{table}[ht]
  \caption{$U_{\mathrm{eff}}$ values (eV) for transition-metal $d$ orbitals. A dash indicates an element not included in the present candidate set (Tc).}
  \label{tab:Uvalues}
  \begin{ruledtabular}
  \begin{tabular}{l cccccccccc}
     & Sc & Ti & V & Cr & Mn & Fe & Co & Ni & Cu & Zn \\
    $U_{\mathrm{eff}}$ & 3 & 3 & 4 & 4 & 4 & 4 & 3 & 7 & 7 & 7 \\[4pt]
     & Y & Zr & Nb & Mo & Tc & Ru & Rh & Pd & Ag & Cd \\
    $U_{\mathrm{eff}}$ & 3 & 3 & 3 & 4 & -- & 4 & 4 & 4 & 0 & 0 \\
  \end{tabular}
  \end{ruledtabular}
\end{table}

\subsection{Spin Splitting Energy}
\label{subsec:sse_quantification}

SSE is the maximum spin-channel energy separation within the occupied valence manifold:
\begin{align}
\mathrm{SSE} &= \max_{\mathbf{k},\,n} \left| E_{n\uparrow}(\mathbf{k}) - E_{n\downarrow}(\mathbf{k}) \right|, \nonumber \\
E_{n\sigma} &\in [E_F - 2~\mathrm{eV},\, E_F].
\label{eq:sse_definition}
\end{align}
Here $n$ denotes the band index as recorded in the VASP \texttt{EIGENVAL} output and $\sigma \in \{\uparrow,\downarrow\}$ labels the spin channel, so that spin-up and spin-down eigenvalues sharing the same index at each $\mathbf{k}$-point are compared directly. The 2~eV window below $E_F$ captures the transport-relevant valence states while avoiding spurious splittings in unoccupied bands. Because SSE reports the largest asymmetry at any single $(\mathbf{k},n)$ pair, it serves as an upper-bound indicator of exchange-driven splitting. That the surrogate nonetheless achieves $R^2 = 0.70$ under grouped cross-validation suggests that the dominant descriptors are robust to the label noise introduced by occasional band-index misassignment.

\subsection{Dataset Generation and Altermagnet Screening}
\label{subsec:dataset_generation}

The initial candidate pool was generated using the MatterGen diffusion-based generative model\cite{zeni2025generative}, which samples from a learned distribution of stable crystal structures and thereby provides structural diversity beyond existing experimental databases. We employed the \texttt{chemical\_system\_energy\_above\_hull} variant, conditioned to minimize predicted energy above the convex hull, generating 100 candidate structures per elemental combination. Magnetic species were selected from the $3d$ and $4d$ transition-metal ranges ($M \in \{\text{Sc--Zn, Y--Cd}\}$) and paired with non-magnetic $p$-block ligands (Supporting Information). These generative outputs served only as initial geometries; all candidates were subsequently refined by the DFT relaxation protocol described above.

A structural pre-filter first retained only candidates containing exactly two magnetic atoms per unit cell, consistent with the two-sublattice collinear setting and subsequent motif-pair construction. The filtered pool was then relaxed via spin-polarized DFT and screened for robust altermagnetic behavior using three joint criteria: (i) near-zero total magnetization ($< 0.005~\mu_B$), ensuring a fully compensated magnetic state; (ii) negligible spin splitting at the $\Gamma$ point (mean value $< 0.01$~eV), excluding materials whose splitting originates at $\Gamma$ rather than at finite momentum; and (iii) significant non-relativistic SSE ($> 0.001$~eV), confirming the presence of momentum-dependent exchange splitting. These criteria identified 323 parent altermagnetic structures, corresponding to approximately 9\% of the filtered pool.

To broaden the structural distribution sampled by the machine-learning model while keeping the chemistry fixed, we applied isotropic and axial strains of 95.0, 97.5, 102.5, and 105.0\% to each parent and performed single-point DFT calculations without further relaxation. Including all successfully evaluated strained variants together with the original parents yielded a final dataset of 3{,}851 entries derived from 323 independent parent chemistries.

\subsection{Motif-Centric Representation}
\label{subsec:motif_representation}

A motif is the local coordination polyhedron formed by the nearest non-magnetic ligands ($X$) surrounding a magnetic center ($M$). The ligand field of each motif governs crystal-field splitting and $p$--$d$ hybridization, which in turn shape the spin-dependent band structure. For altermagnetic materials, the two antiparallel magnetic sublattices occupy site-inequivalent coordination environments, breaking the degeneracy-protecting symmetries present in conventional collinear antiferromagnets and enabling momentum-dependent exchange splitting\cite{vsmejkal2022beyond,vsmejkal2022emerging}. Motifs were extracted from DFT-relaxed structures using the CrystalNN algorithm\cite{pan2021benchmarking} as implemented in \textit{pymatgen}\cite{ong2013python}. For each magnetic site $i$, the motif $\mathcal{M}_i$ consists of the relative ligand coordinates,
\begin{equation}
\mathcal{M}_i = \{\mathbf{r}_{X_1}, \ldots, \mathbf{r}_{X_N}\},
\quad
\mathbf{r}_{X_j} \equiv \mathbf{R}_{X_j}-\mathbf{R}_{M},
\label{eq:motif_definition}
\end{equation}
where $N$ is the coordination number. Because each structure contains exactly two magnetic atoms per unit cell, every structure yields a paired set $(\mathcal{M}_1,\mathcal{M}_2)$ associated with antiparallel sublattices. All descriptors are computed entirely from atomic coordinates and elemental identities; no electronic-structure-derived quantities enter the feature space.

\subsection{Structural and Compositional Descriptors}
\label{subsec:descriptors}

Three primary descriptors encode the symmetry, geometry, and chemistry governing altermagnetic splitting: MSBI captures motif-level coordination inequivalence, MPF quantifies spatial packing of the ligand cage, and the $p/d$ electron ratio provides a proxy for hybridization propensity. SHAP analysis identifies these three as the dominant predictors of SSE. The complete set of additional descriptors is listed in Supporting Information, Table~S1.

\subsubsection{Motif Symmetry Breaking Index (MSBI)}
\label{subsubsec:methods_msbi}

Altermagnetism requires the absence of degeneracy-protecting symmetry operations that map one magnetic sublattice onto the other while preserving the crystal potential\cite{vsmejkal2022beyond,vsmejkal2022emerging}. In collinear antiferromagnets, spin degeneracy is protected by $\mathcal{PT}$ symmetry, which requires the spatial environments of the two sublattices to be related by either identity or inversion. Microscopically, this condition is violated when the two antiparallel magnetic sites experience locally inequivalent coordination environments---for instance, rotated or non-superimposable ligand cages. Because our dataset is restricted to binary compounds, where both magnetic sites are coordinated by the same ligand species, this inequivalence can be assessed through a purely geometric comparison of the paired motifs.

To convert this discrete symmetry criterion into a continuous design variable, we introduce the Motif Symmetry-Breaking Index (MSBI). For a given structure, the two magnetic motifs are represented as ligand-vector sets $\mathcal{A}=\{\mathbf{a}_1,\ldots,\mathbf{a}_N\}$ and $\mathcal{B}=\{\mathbf{b}_1,\ldots,\mathbf{b}_N\}$ relative to their respective magnetic centers, with coordination numbers determined by CrystalNN\cite{pan2021benchmarking,ong2013python}. When the two sites yield different coordination counts ($N_1\neq N_2$), a common cardinality $N=\min(N_1,N_2)$ is enforced by retaining the $N$ neighbors with the largest CrystalNN weights. All comparisons are performed in the fixed crystal coordinate frame, without Kabsch-type rotational alignment. This choice is deliberate: Kabsch alignment would artificially restore the rotational degree of freedom that the crystal itself has broken, erasing precisely the geometric information we seek to capture. Because the relative orientation of the two motifs is fixed by the crystal symmetry operations connecting them, any nonzero relative rotation is itself a signature of broken $\mathcal{PT}$ and must enter the metric explicitly. Deviations are rendered dimensionless by the mean metal--ligand bond length,
\begin{equation}
l_0 = \frac{1}{2}\left(\frac{1}{N}\sum_{i=1}^{N}\|\mathbf{a}_i\|+\frac{1}{N}\sum_{i=1}^{N}\|\mathbf{b}_i\|\right),
\label{eq:l0}
\end{equation}
ensuring comparability across structures with different absolute bond lengths.

\paragraph{Motif Direct Deviation ($D_d$).}
The dissimilarity between $\mathcal{A}$ and $\mathcal{B}$ is quantified via a permutation-invariant RMSD that optimizes over all one-to-one ligand pairings,
\begin{equation}
D_d(\mathcal{A}, \mathcal{B}) = \frac{1}{l_0} \min_{\tau \in S_N} \left( \frac{1}{N} \sum_{i=1}^{N} \left\| \mathbf{a}_{\tau(i)} - \mathbf{b}_i \right\|^2 \right)^{1/2},
\label{eq:Dd}
\end{equation}
where $S_N$ denotes the set of all $N!$ one-to-one pairings between ligands of $\mathcal{A}$ and $\mathcal{B}$, and $\tau$ is the specific pairing that minimizes the summed squared distance. In practice, this combinatorial optimization is solved efficiently as a linear-sum assignment problem via the Hungarian algorithm\cite{kuhn1955hungarian}. A small $D_d$ indicates that the two motifs are nearly identical in the crystal frame, consistent with a $\mathcal{PT}$-protecting identity relation.

\paragraph{Motif Inversion Deviation ($D_I$).}
To test whether the motifs are instead related by spatial inversion, we compare the point-inverted set $-\mathcal{A}=\{-\mathbf{a}\mid\mathbf{a}\in\mathcal{A}\}$ with $\mathcal{B}$ using the same permutation-invariant metric:
\begin{multline}
D_I(\mathcal{A}, \mathcal{B}) \equiv D_d(-\mathcal{A},\, \mathcal{B}) \\
= \frac{1}{l_0} \min_{\tau \in S_N} \left( \frac{1}{N} \sum_{i=1}^{N} \left\| -\mathbf{a}_{\tau(i)} - \mathbf{b}_i \right\|^2 \right)^{1/2}.
\label{eq:DI}
\end{multline}
A small $D_I$ signals that the two motifs are close to inversion-related, the remaining $\mathcal{PT}$-protecting configuration.

\paragraph{MSBI.}
The index is defined as the product of the two deviations,
\begin{equation}
\mathrm{MSBI} \equiv D_d(\mathcal{A}, \mathcal{B}) \times D_I(\mathcal{A}, \mathcal{B}).
\label{eq:MSBI}
\end{equation}
Because $D_d$ and $D_I$ quantify the deviation from each of the two spatial relations that can sustain $\mathcal{PT}$ symmetry, a large MSBI directly signals $\mathcal{PT}$ symmetry breaking. The multiplicative form ensures that MSBI vanishes whenever either relation is approximately satisfied, while structures in which both are strongly broken receive amplified values, sharpening the distinction between marginal and robust altermagnet candidates. MSBI serves as a continuous design variable during descriptor construction, SHAP-based ranking, and surrogate training; during Bayesian optimization, it enters as an empirically bounded proxy for reasons detailed in Section~\ref{subsec:inverse_design}.

\subsubsection{Motif Packing Fraction (MPF)}
\label{subsubsec:methods_mpf}

MPF measures the fractional unit-cell space occupied by the ligand cage. The motif dimensionality is determined by SVD of the mean-centered ligand coordinates: a motif is classified as planar ($D_{\mathrm{motif}}=2$) when the singular-value spectrum has effective rank two, and as polyhedral ($D_{\mathrm{motif}}=3$) otherwise. For three-dimensional motifs,
\begin{equation}
\mathrm{MPF} = \frac{V_{\mathrm{motif}}}{V_{\mathrm{cell}}},
\label{eq:mpf_3d}
\end{equation}
where $V_{\mathrm{motif}}$ is the convex-hull volume of the ligand set. For planar motifs (e.g., square-planar coordination), an area-based analog is used:
\begin{equation}
\mathrm{MPF} = \frac{A_{\mathrm{motif}}}{A_{\mathrm{cell}}},
\label{eq:mpf_2d}
\end{equation}
where $A_{\mathrm{motif}}$ is the convex-hull area of the ligand set projected onto its best-fit plane and $A_{\mathrm{cell}}$ is the area of the unit-cell face whose normal is most closely aligned with the motif-plane normal. Planar motifs constitute only $2.2\%$ of the parent prototype pool; a feature-ablation study separating 2D and 3D MPF into independent columns leaves the SHAP ranking of the three dominant descriptors unchanged (Section~S4). Larger MPF values correspond to more extended ligand cages relative to the cell, providing a geometric proxy for packing density.

\subsubsection{$p/d$ Electron Ratio}
\label{subsubsec:methods_pd_ratio}

The $p/d$ electron ratio,
\begin{equation}
p/d \; \text{electron ratio} = \frac{n_{p,X}}{n_{d,M}},
\label{eq:pd_ratio}
\end{equation}
uses valence electron counts from the PBE pseudopotential configurations employed in the DFT calculations ($n_{p,X}$ for the ligand $p$ shell, $n_{d,M}$ for the metal $d$ shell; see Table~S2 for numerical values) as a composition-level proxy for $p$--$d$ hybridization propensity. Because the pseudopotential-derived count can deviate from in-crystal occupation in strongly charge-transferred systems\cite{zaanen1985band}, the ratio is best viewed as a coarse screening variable---a trade-off that preserves DFT-free applicability. Compositions with $p/d < 1$---where the metal $d$ shell is electron-rich relative to the ligand $p$ shell---tend to exhibit stronger $p$--$d$ hybridization and, correspondingly, larger SSE.

\subsection{Machine-Learning Workflow}
\label{subsec:ml_workflow}

XGBoost\cite{chen2016xgboost} was trained on \texttt{log1p}-transformed SSE using the 52 motif-based descriptors (Supporting Information, Table~S1). Generalization was assessed via 10-fold grouped cross-validation, with folds partitioned by parent structure so that strain-derived variants never leak between training and evaluation sets. The surrogate achieves $R^2 = 0.70$ and MAE $= 123.1$~meV across held-out folds, sufficient for candidate ranking prior to DFT validation. SHAP analysis\cite{lundberg2017unified}, averaged over 100 independently seeded models, identifies MSBI, MPF, and the $p/d$ electron ratio as the three dominant predictors of SSE. Full details of the cross-validation protocol, hyperparameter optimization, and SHAP rescaling procedure are provided in the Supporting Information.

\subsection{Inverse Design via Bayesian Optimization}
\label{subsec:inverse_design}

Bayesian optimization (BO) was coupled to the trained XGBoost surrogate to search for candidate materials with large predicted spin-splitting energy (SSE). Each BO trial proposes a design vector $\boldsymbol{\theta}$, which is mapped to the surrogate input $\mathbf{x}_{\mathrm{BO}}(\boldsymbol{\theta})$ through three complementary routes. First, the composition $(Z_M, Z_X)$ fixes the electronic descriptors deterministically. Second, a virtual motif is initialized from an ideal tetrahedral, square-planar, or octahedral template and perturbed via radial and angular distortion parameters in spherical coordinates, yielding a concrete set of ligand coordinates from which all intra-motif structural descriptors are extracted algorithmically. Third, global quantities that cannot be constructed from a single virtual motif---including MSBI and $\sigma_{\mathrm{inhom}}$, which are inter-motif quantities requiring two distinct coordination environments, as well as inter-motif distances and $V_{\mathrm{cell}}$---are sampled within the empirical bounds of the reference dataset, confining the search to a chemically informed region of descriptor space.

The surrogate objective is
\begin{equation}
\boldsymbol{\theta}^{\ast} = \arg\max_{\boldsymbol{\theta}}\; f_{\mathrm{XGB}}\!\left(\mathbf{x}_{\mathrm{BO}}(\boldsymbol{\theta})\right),
\label{eq:bo_sse_objective}
\end{equation}
where the search domain is bounded by the empirical ranges of the training data. The tree-structured Parzen estimator (TPE) algorithm\cite{bergstra2011algorithms}, as implemented in Optuna, navigates this mixed discrete--continuous space over $10^{5}$ trials. Because MSBI is treated as a sampled global proxy rather than a directly constructed geometric descriptor, the optimizer learns which MSBI ranges are associated with large predicted SSE instead of enforcing a fixed target value. Top candidates are stratified by coordination class ($N_X = 4$ for square-planar and $N_X = 6$ for octahedral geometries) so that chemically distinct ligand-field environments are evaluated separately during ranking.

For DFT validation, each optimized descriptor vector is matched to the nearest structural prototype in the reference dataset. Matching is performed in $z$-score-normalized structural-descriptor space using lexicographic ranking: cosine similarity is maximized first to identify prototypes sharing the same structural archetype, and Euclidean distance serves as a tiebreaker to select the closest match in absolute feature magnitude. The prototype species are then replaced with the BO-identified elements, and the resulting periodic structure is independently relaxed with spin-polarized DFT. Full details of the search-space parameterization, prototype-matching procedure, and convergence behavior are provided in the Supporting Information.


\section{Results and Discussion}
\label{sec:results}

\subsection{Three Descriptors Govern Altermagnetic Splitting}

The XGBoost surrogate trained on the 3,851-entry altermagnetic dataset achieves $R^2 = 0.70$ and a mean absolute error of 123.1~meV in grouped cross-validation, using only structurally and compositionally derived descriptors without DFT-derived input features. This performance confirms that the 52-descriptor feature space carries sufficient information to reproduce DFT-calculated SSE with adequate predictive power for screening.

To identify which descriptors drive these predictions, we analyzed the trained model using SHAP\cite{lundberg2017unified}, which assigns each descriptor a signed contribution to every individual prediction. For each descriptor $i$, the mean absolute SHAP magnitude,
\begin{equation}
\bar{S}_i \;=\; \frac{1}{N}\sum_{j=1}^{N} \bigl|s_{ij}\bigr|,
\end{equation}
where $s_{ij}$ is the SHAP value for feature $i$ in structure $j$, quantifies global importance. All SHAP statistics reported here are averaged over 100 independently trained models to ensure robustness against stochastic variations in tree-based learning\cite{chen2016xgboost}.

\begin{figure*}[ht]
    \centering
    \includegraphics[width=\textwidth]{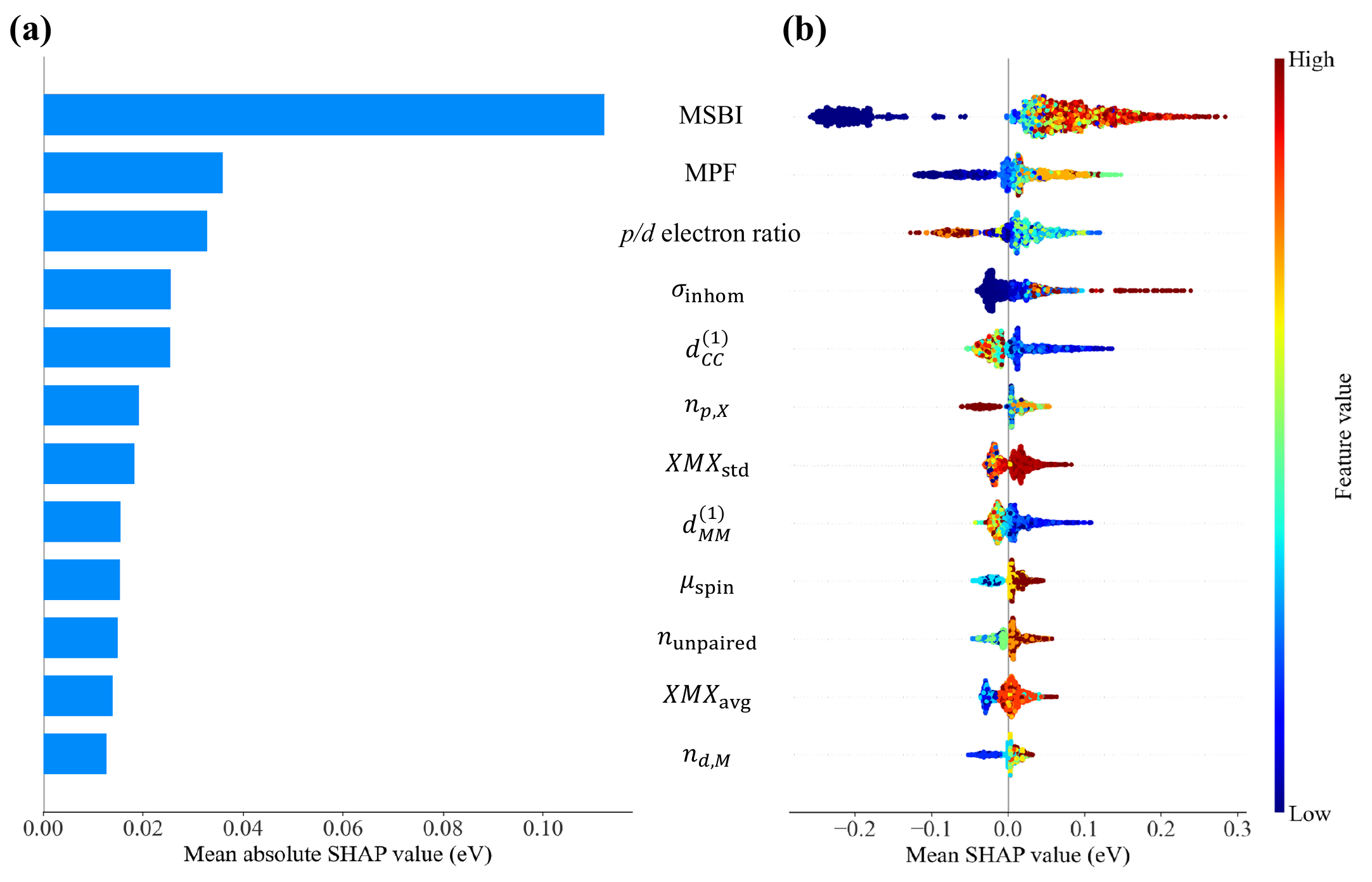}
    \caption{Three descriptors dominate SSE prediction.
    (a) Top-12 descriptors ranked by mean absolute SHAP value $\bar{S}_i$ (eV), averaged over 100 independently trained XGBoost models. MSBI, MPF, and the $p/d$ ratio together account for 37.4\% of total SHAP attribution. Feature labels are shared with panel~(b).
    (b) SHAP beeswarm plot for the same 12 descriptors. Each point represents one structure; horizontal position shows the signed SHAP value (eV). Color indicates normalized feature value (red = high, blue = low). Positive SHAP values indicate that the feature increases predicted SSE.
    }
    \label{fig:shap}
\end{figure*}

\begin{table}[ht]
\caption{Top-12 descriptors ranked by mean absolute SHAP value $\bar{S}_i$ (eV). The normalized contribution ratio $R_i = \bar{S}_i / \sum_k \bar{S}_k$ expresses each descriptor's share of the model's total explanatory budget; the cumulative fraction is the running sum of $R_i$. Both are computed relative to the total SHAP magnitude ($\sum_k \bar{S}_k = 0.4836$~eV). The top three descriptors account for 37.4\% of the model's total attribution.
}
\label{tab:shap_top12}
\begin{ruledtabular}
\begin{tabular}{lccc}
Feature (Symbol) & Mean $|\text{SHAP}|$ (eV) & Ratio $R_i$ & Cumulative \\
\hline
MSBI & 0.1122 & 0.2320 & 0.2320 \\
MPF & 0.0359 & 0.0743 & 0.3063 \\
$p/d$ electron ratio & 0.0328 & 0.0679 & 0.3742 \\
$\sigma_{\text{inhom}}$ & 0.0256 & 0.0529 & 0.4270 \\
$d_{CC}^{(1)}$ & 0.0255 & 0.0526 & 0.4797 \\
$n_{p,X}$ & 0.0191 & 0.0395 & 0.5192 \\
$XMX_{\text{std}}$ & 0.0183 & 0.0378 & 0.5570 \\
$d_{MM}^{(1)}$ & 0.0154 & 0.0319 & 0.5890 \\
$\mu_{\text{spin}}$ & 0.0154 & 0.0318 & 0.6208 \\
$n_{\text{unpaired}}$ & 0.0149 & 0.0309 & 0.6516 \\
$XMX_{\text{avg}}$ & 0.0140 & 0.0290 & 0.6806 \\
$n_{d,M}$ & 0.0127 & 0.0262 & 0.7069 \\
\end{tabular}
\end{ruledtabular}
\end{table}

\textbf{Figure~2}(a) reveals a pronounced hierarchy among the 52 descriptors. The Motif Symmetry-Breaking Index (MSBI) dominates with $\bar{S}_{\mathrm{MSBI}} = 0.1122$~eV, accounting for $R_{\mathrm{MSBI}} = 23.2\%$ of the total SHAP attribution (total $\sum_k \bar{S}_k = 0.4836$~eV)---more than three times the contribution of any other single feature. The motif packing fraction (MPF, $R = 7.4\%$) and the $p/d$ electron ratio ($R = 6.8\%$) complete the top three, which together capture 37.4\% of the model's explanatory budget. \textbf{Table~2} lists the top-12 descriptors; the cumulative contribution reaches 70.7\% by rank~12, but the rapid falloff beyond the top three indicates that the essential design space is low-dimensional. The remaining 40 descriptors collectively account for less than 30\% of the total attribution, confirming that SSE is governed by a compact set of physically interpretable variables.

The beeswarm plot in Figure~2(b) reveals the directional structure underlying these importances. MSBI shows an unambiguous positive trend: larger motif-level symmetry breaking systematically increases SSE, consistent with its role as a continuous measure of $\mathcal{PT}$ symmetry breaking between antiparallel sublattices\cite{vsmejkal2022emerging,vsmejkal2022beyond}. MPF exhibits a similar positive correlation, where denser packing drives predictions upward through enhanced superexchange coupling from increased inter-site hopping\cite{anderson1959new,coey2010magnetism}. The $p/d$ ratio displays the opposite behavior: $p/d < 1$, where the metal $d$ shell dominates, favors strong $p$--$d$ hybridization and larger SSE, while $p/d > 1$ suppresses splitting\cite{zaanen1985band}. Crucially, all three leading descriptors depend on experimentally controllable variables---lattice distortions, applied pressure, and chemical substitution---enabling their direct use as optimization targets in inverse design. The following subsections examine each mechanism in detail and validate the resulting design rules with DFT calculations.

\subsection{MSBI Encodes Continuous Symmetry Breaking}

The existence of spin-split bands in altermagnets originates from a specific crystal symmetry requirement: the magnetic sublattices must be connected by a crystal rotation or mirror operation, distinct from spatial inversion ($\mathcal{P}$) or simple translation\cite{vsmejkal2022emerging,vsmejkal2020crystal}. The nature and magnitude of this effective $\mathcal{PT}$ symmetry breaking determine the size of SSE. To quantify this non-relativistic effect directly from the crystal structure, we introduced the Motif Symmetry-Breaking Index (MSBI), defined as the product $D_d \times D_I$ of two geometric deviations. The multiplicative form ensures that MSBI remains small whenever either degeneracy-protecting relation is approximately satisfied---near-identity of the two motifs ($D_d \approx 0$) or near-inversion partnership ($D_I \approx 0$)---and becomes large only when both are absent, signaling that the combined $\mathcal{PT}$ symmetry enforcing Kramers-like spin degeneracy in conventional antiferromagnets is broken\cite{hayami2019momentum}. MSBI thus provides a single continuous scalar that captures the essential structural condition for non-relativistic spin splitting identified by group-theoretic analyses\cite{cheong2025altermagnetism}.

\begin{figure*}[ht!]
    \centering
    \includegraphics[width=\textwidth]{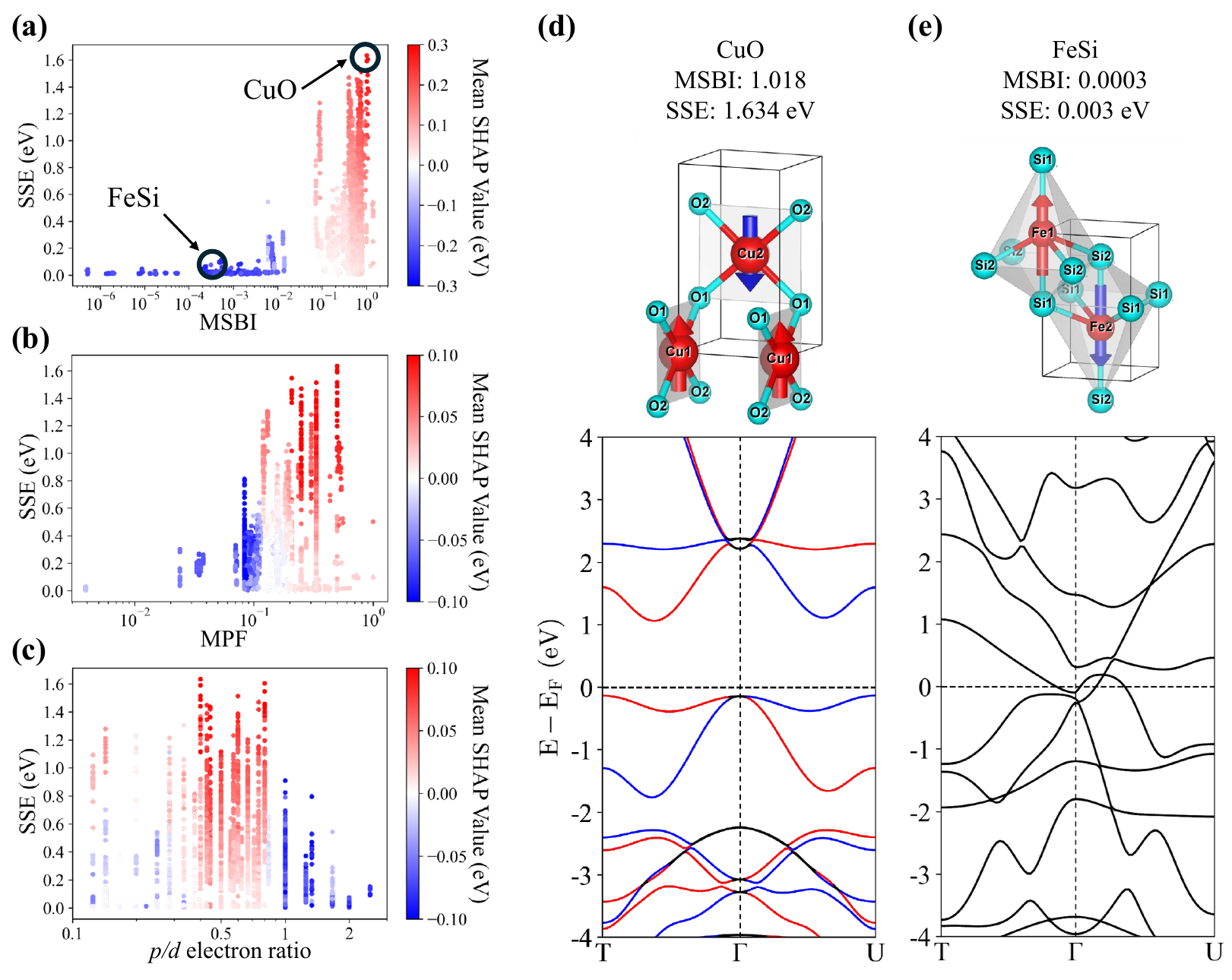}
    \caption{MSBI controls symmetry breaking and SSE.
    (a--c) SSE versus (a) MSBI, (b) MPF, and (c) $p/d$ electron ratio, with points colored by mean SHAP value (eV) for the corresponding descriptor. All three panels use logarithmic $x$-axes. MSBI exhibits a sharp threshold near 0.5, above which SSE systematically exceeds 0.4~eV. CuO and FeSi are annotated in (a) as extreme-MSBI case studies.
    (d) CuO (MSBI = 1.018, SSE = 1.634~eV): crystal structure with magnetic atoms (red) and ligands (cyan), motif projection showing coordination geometry and spin orientations, and spin-resolved band structure near $E_F$ (red = spin-up, blue = spin-down).
    (e) FeSi (MSBI = 0.0003, SSE = 0.003~eV): crystal structure, motif projection, and band structure. Bands are plotted in black as spin-up and spin-down channels are visually indistinguishable.
    }
    \label{fig:msbi}
\end{figure*}

Figure~3(a--c) quantifies the descriptor--SSE relationships across the full dataset. Panel~(a) reveals a sharp design threshold for MSBI: structures with MSBI $> 0.5$ systematically yield SSE $> 0.4$~eV, while MSBI $< 0.1$ produces negligible splitting regardless of other favorable properties. The SHAP coloring reinforces this trend---positive SHAP contributions (red) concentrate overwhelmingly in the high-MSBI regime. Panel~(b) shows that the largest splittings (SSE $> 0.8$~eV) require MPF $> 0.20$, while panel~(c) confirms that moderate $p/d$ values ($0.4 \leq p/d \leq 1$) span the full SSE range, whereas $p/d \ge 1$ confines SSE below approximately 0.6~eV. Together, these three panels establish quantitative design boundaries---MSBI $> 0.5$, MPF $> 0.20$, $p/d < 1$---that jointly define the favorable region in descriptor space; the mechanistic origins of the MPF and $p/d$ trends are developed in the following subsection.

To illustrate the physical content of the MSBI threshold, panels~(d) and~(e) contrast two structures with extreme MSBI values. In the CuO-type structure (Figure~3d), the two Cu sublattices experience strongly inequivalent square-planar coordination environments (MSBI = 1.018), and the resulting motif-level asymmetry produces pronounced spin-up/spin-down band separation with SSE = 1.634~eV. In contrast, FeSi (Figure~3e) crystallizes in a structure where the two magnetic sublattices are nearly perfect inversion partners ($D_I \approx 0$), resulting in a negligible MSBI of 0.0003. In this configuration, the combined $\mathcal{PT}$ symmetry remains effectively intact, enforcing Kramers-like spin degeneracy across the Brillouin zone. Accordingly, the band structure shows perfect spin degeneracy within numerical precision (SSE = 3~meV). The contrast between these two structures---from negligible splitting (3~meV) to giant SSE (1.634~eV) across three orders of magnitude in MSBI---underscores the descriptor's quantitative predictive power within the collinear two-sublattice setting.

The CuO/FeSi contrast also highlights the key conceptual advance that MSBI provides over existing classification schemes. Magnetic point group (MPG) analysis rigorously identifies the symmetry operations connecting opposite-spin sublattices and classifies materials into altermagnetic symmetry classes\cite{cheong2025altermagnetism,radaelli2024tensorial}, but it is inherently discrete and binary: a material either belongs to an altermagnetic class or it does not. MPG analysis can establish whether altermagnetic splitting is symmetry-allowed; it cannot predict how large the splitting will be. MSBI complements this framework by providing a continuous scalar that quantifies the magnitude of local coordination inequivalence and correlates directly with the achievable SSE. This continuity is not a cosmetic improvement. It is the operation that promotes altermagnet design from a binary classification problem into a well-posed optimization problem on which surrogate-based search and gradient-based methods can act---precisely what the SHAP hierarchy of Figure~2 and the sharp threshold of Figure~3(a) make quantitatively visible.

\subsection{Packing and Covalency Amplify Exchange-Driven Splitting}

The MSBI threshold establishes the structural prerequisite for altermagnetic splitting, but the magnitude of SSE depends on two additional factors that sit downstream of symmetry breaking: how tightly motifs pack in the lattice (MPF) and the balance between ligand and metal valence electrons ($p/d$ ratio). These descriptors capture, respectively, the energy scale of exchange interactions and the strength of the covalent mixing that mediates spin-dependent band splitting. Because both mechanisms operate on top of a symmetry-breaking baseline already set by MSBI, they explain why many structures with permissive symmetry (moderate MSBI) nonetheless exhibit negligible SSE---and how targeted chemical design can activate large splitting within a given structural framework.

\begin{figure*}[ht!]
    \centering
    \includegraphics[width=\textwidth]{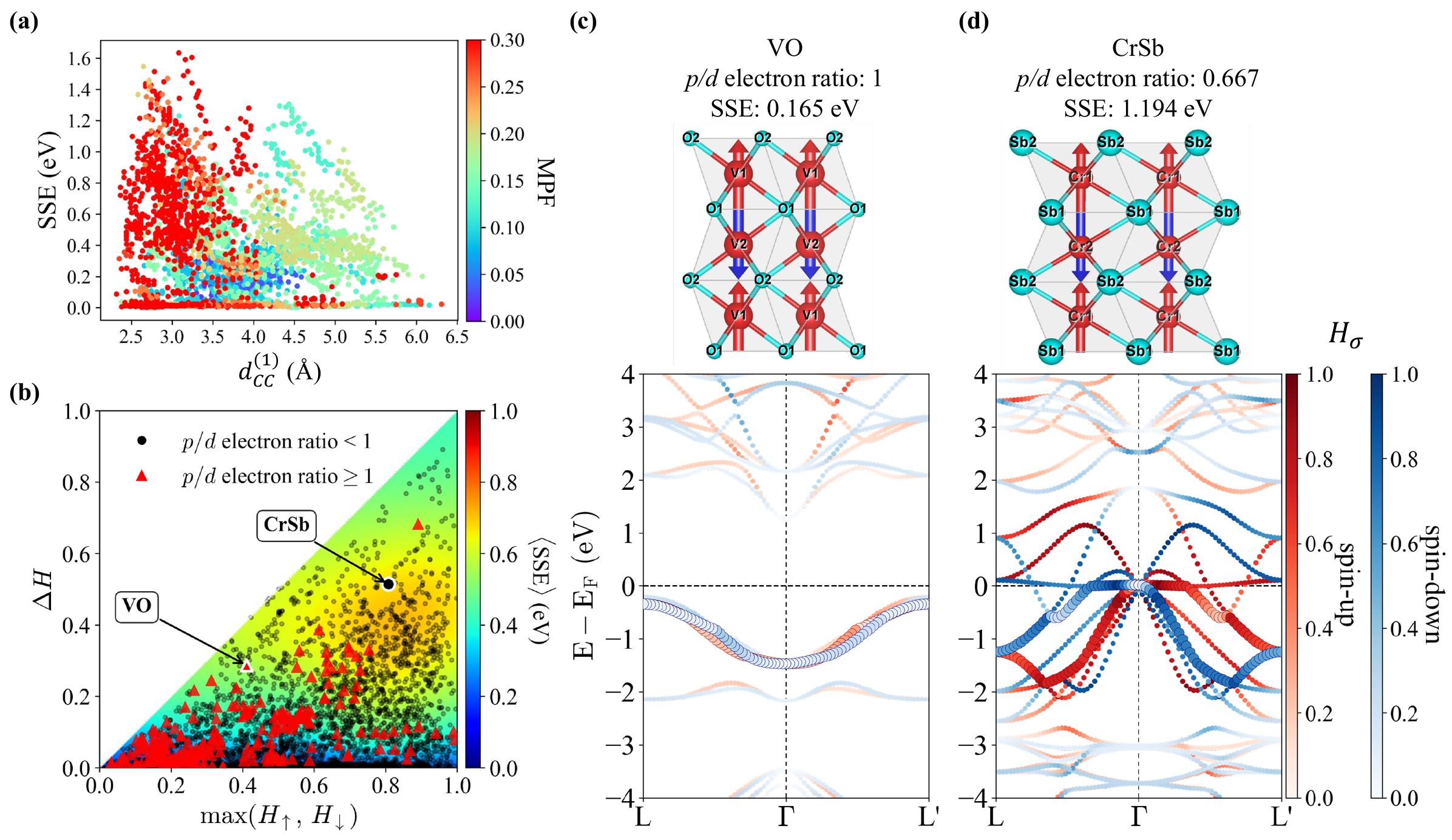}
    \caption{Packing and covalency jointly govern SSE magnitude.
    (a) SSE versus nearest motif-centroid distance $d_{CC}^{(1)}$, colored by MPF. Large SSE ($>0.8$~eV) requires high packing (MPF $> 0.20$) and short inter-motif separations ($d_{CC}^{(1)} < 3.5$~\AA).
    (b) Hybridization spin asymmetry $\Delta H = |H_{\uparrow} - H_{\downarrow}|$ versus maximum spin-resolved $p$--$d$ hybridization $\max(H_{\uparrow}, H_{\downarrow})$, colored by SSE. Black circles denote $p/d < 1$; red triangles denote $p/d \geq 1$. Large SSE concentrates in the upper-right region (strong hybridization and large spin asymmetry, predominantly $p/d < 1$). VO and CrSb are annotated as case studies.
    (c) VO ($p/d = 1$, SSE = 0.165~eV): crystal structure in the P$6_3$/mmc prototype with spin orientations indicated (top), and spin-resolved $p$--$d$ hybridization fatbands along $L$--$\Gamma$--$L'$ (bottom), showing weak, spin-symmetric hybridization.
    (d) CrSb ($p/d = 0.667$, SSE = 1.194~eV): crystal structure (top) and fatbands (bottom), displaying strong, spin-asymmetric hybridization. In (c) and (d), band opacity scales with $H_\sigma$; separate colorbars indicate spin-up (red) and spin-down (blue) channels. The band pair for which SSE is evaluated is highlighted with a thicker linewidth.
    }
    \label{fig:packing}
\end{figure*}

Figure~4(a) reveals the role of packing density. SSE is plotted against the nearest motif-centroid distance $d_{CC}^{(1)}$, with MPF encoded by color. The largest splittings (SSE $> 0.8$~eV) cluster exclusively in a compact region characterized by high packing fractions (MPF $> 0.20$) and short inter-motif separations ($d_{CC}^{(1)} < 3.5$~\AA), while structures with MPF $< 0.15$ are confined below SSE $\approx 0.4$~eV. The underlying mechanism is exchange enhancement through increased orbital overlap. In the superexchange picture, the exchange coupling scales as $J \sim t^2/U$, where $t$ is the hopping integral and $U$ the on-site Coulomb repulsion\cite{anderson1959new,coey2010magnetism}. Because $t$ decays approximately exponentially with interatomic distance, $J$ diminishes rapidly as motifs separate, accounting for the systematic confinement of low-MPF structures to the small-SSE regime. However, Figure~4(a) also contains high-MPF structures with negligible SSE, demonstrating that dense packing is necessary but insufficient: favorable exchange geometry must be accompanied by appropriate symmetry breaking (high MSBI) and electronic structure (favorable $p/d$ ratio). The three descriptors are thus complementary, not redundant.

To isolate the covalency axis from the geometric ones, we exploit a natural controlled experiment within the dataset. VO and CrSb both crystallize in the P$6_3$/mmc prototype and have nearly identical MSBI, so that structural symmetry breaking and packing are effectively fixed and composition is the sole variable. VO has $p/d = 1$ and SSE = 0.165~eV; CrSb has $p/d = 0.667$ and SSE = 1.194~eV---a sevenfold enhancement driven entirely by compositional substitution within a single host lattice. The dataset-wide trend behind this comparison is that SSE drops below approximately 0.6~eV once $p/d$ exceeds unity, revealing a strong nonlinear dependence rationalized within the Zaanen--Sawatzky--Allen (ZSA) classification scheme\cite{zaanen1985band}. The charge-transfer energy $\Delta_{\mathrm{CT}} = |\epsilon_d - \epsilon_p|$ controls the degree of $p$--$d$ hybridization: larger $\Delta_{\mathrm{CT}}$ suppresses covalent mixing, reducing the effective covalency and ultimately the SSE. Halide ligands and early transition metals---which tend toward the Mott--Hubbard limit with large $\Delta_{\mathrm{CT}}$\cite{bocquet1992electronic,imada1998metal}---systematically populate the high-$p/d$, low-SSE regime, while late $3d$ metals paired with chalcogenide or pnictide ligands occupy the charge-transfer regime where strong covalency enhances exchange-driven splitting.

To connect this compositional rule to electronic structure, we define a spin-resolved $p$--$d$ hybridization index for each spin channel $\sigma \in \{\uparrow, \downarrow\}$,
\begin{equation}
  H_\sigma = 2 \cdot \min\!\left( p_{\mathrm{global},\sigma},\, d_{\mathrm{global},\sigma} \right),
\end{equation}
where $p_{\mathrm{global},\sigma}$ and $d_{\mathrm{global},\sigma}$ are the normalized ligand-$p$ and metal-$d$ projected orbital weights at the $k$-point and band where the SSE is evaluated, computed as
\begin{align}
p_{\mathrm{global},\sigma}
&= \frac{p_\sigma}{s_\sigma + p_\sigma + d_\sigma}, \nonumber \\
d_{\mathrm{global},\sigma}
&= \frac{d_\sigma}{s_\sigma + p_\sigma + d_\sigma},
\end{align}
with $s_\sigma$, $p_\sigma$, and $d_\sigma$ denoting the spin-resolved orbital contributions. By construction, $H_\sigma$ reaches unity for ideal 1:1 $p$--$d$ mixing and is used only as a diagnostic of the DFT band structure, not as an input descriptor. Figure~4(b) maps the relationship between hybridization strength, spin asymmetry, and SSE. The SSE magnitude increases systematically from the lower-left corner (weak hybridization, small spin asymmetry) toward the upper-right region (strong hybridization, large spin asymmetry). Structures with $p/d < 1$ (black circles) dominate the high-SSE region, while those with $p/d \geq 1$ (red triangles) cluster near the low-SSE baseline. Strong covalency ($H_\sigma$ large) establishes a favorable baseline, while a pronounced spin-channel imbalance ($\Delta H$ large) further amplifies splitting.

Panels~(c) and~(d) of Figure~4 demonstrate this mechanism through the VO/CrSb controlled comparison. In VO (panel~c), the ratio $p/d \approx 1$ places the system in a regime where the charge-transfer energy suppresses $p$--$d$ covalency: the bands near $E_F$ are uniformly faint ($H_\sigma \approx 0$) and approximately spin-symmetric, producing minimal splitting. In CrSb (panel~d), the reduced $p/d$ ratio activates the covalency mechanism, producing strong overall hybridization (fat bands near $E_F$) and pronounced spin asymmetry, where a strongly hybridized band in one spin channel is paired with a weakly hybridized counterpart in the opposite channel. The VO/CrSb comparison thus validates the design rule within a single structural prototype and demonstrates that compositional tuning alone---without any change in crystal structure---can activate giant altermagnetic splitting in an otherwise modest host lattice.

\subsection{Bayesian Optimization Validates the Three-Axis Framework}

To test whether the descriptor-based rules can guide candidate generation, we applied the BO workflow to the trained XGBoost surrogate. Over $10^{5}$ TPE trials\cite{bergstra2011algorithms}, the optimizer sampled compositions and motif geometries, assembled the 52-descriptor input vector from deterministic descriptor evaluations and empirically bounded global proxies, and maximized the predicted SSE. \textbf{Table~3} summarizes the highest-ranked candidates together with their independent DFT-validated SSE values.

\begin{table*}[ht]
    \centering
    \caption{Top BO candidates for each coordination class, their nearest structural prototypes, and independent DFT validation results. Structural similarity is quantified by cosine similarity $c$ and Euclidean distance $d$ in the normalized structural-descriptor space. For reference, the median pairwise Euclidean distance among the 323 parent prototypes in the dataset is $d \approx 7.8$ in this space.}
    \label{tab:nn_motifs}
    \begin{ruledtabular}
        \begin{tabular}{cccccccc}
            Coordination ($N_X$) & Rank & $M$--$X$ composition & Prototype & Pred.\ SSE (eV) & DFT SSE (eV) & Cos.\ sim.\ $c$ & Distance $d$ \\
            \hline
            4 & 1 & Fe--S$^{\mathrm{a}}$    & Square-planar CuO  & 1.061 & 1.297 & 0.694 &  9.17 \\
            4 & 3 & Ni--S$^{\mathrm{b}}$    & Square-planar CuO  & 1.042 & --    & 0.555 & 12.93 \\
            4 & 4 & Co--O        & Square-planar CuO  & 1.033 & 0.972 & 0.634 & 10.95 \\
            4 & 5 & Cr--S        & Square-planar CuO  & 1.009 & 0.552 & 0.662 &  9.75 \\
            \hline
            6 & 1 & Ni--S        & NiAs               & 0.893 & 0.823 & 0.897 & 20.72 \\
            6 & 2 & Co--S        & NiAs               & 0.874 & 1.103 & 0.905 & 24.35 \\
            6 & 3 & Cu--S$^{\mathrm{c}}$    & NiAs               & 0.868 & --    & 0.904 & 21.63 \\
            6 & 4 & Fe--As       & NiAs               & 0.867 & 1.089 & 0.893 & 21.00 \\
        \end{tabular}
    \end{ruledtabular}
    \begin{flushleft}
    \footnotesize{$^{\mathrm{a}}$ Another Fe--S motif appears at rank 2 and relaxes to the same ground-state structure.}\\
    \footnotesize{$^{\mathrm{b}}$ DFT relaxation yields a non-magnetic ground state in square-planar coordination; octahedral Ni--S remains magnetically stable and ranks first for $N_X = 6$.}\\
    \footnotesize{$^{\mathrm{c}}$ Cu--S candidates relax to non-magnetic states, consistent with the closed-shell $d^{10}$ configuration of Cu$^+$.}
    \end{flushleft}
\end{table*}

Table~3 shows a clear separation by coordination class. For $N_X = 4$, the optimizer concentrates on square-planar CuO-type environments with predicted SSE values of 1.01--1.06~eV. For $N_X = 6$, it converges on NiAs-type motifs with predicted SSE of 0.85--0.89~eV and high cosine similarity ($c \approx 0.89$--$0.91$). Across both classes, mid-to-late $3d$ metals paired with chalcogenide or pnictide ligands dominate, consistent with the descriptor trends identified from the training set.

DFT screening excludes two of the eight top-ranked candidates (75\% magnetic-stability retention rate) for chemically transparent reasons: Cu--S relaxes to a non-magnetic closed-shell $d^{10}$ state, and square-planar Ni--S collapses to a low-spin $S = 0$ configuration. These exclusions show that the SSE-only surrogate correctly identifies the symmetry-breaking and covalency ingredients of large spin splitting but is deliberately blind to magnetic stability. Because DFT-computed atomic magnetic moments were not included in the descriptor set---they would violate the DFT-free constraint---a single-shot coordination-aware spin-state prescreen applied after BO would remove such false positives at negligible cost.

Of the six magnetically stable candidates, five yield DFT-validated SSE exceeding 0.8~eV. The octahedral candidates (Co--S: 1.103~eV, Fe--As: 1.089~eV, Ni--S: 0.823~eV) all exceed or closely match their surrogate predictions, indicating conservative behavior in this well-represented coordination class. For the square-planar candidates, Fe--S (1.297~eV) and Co--O (0.972~eV) validate strongly, while Cr--S (0.552~eV) is substantially overpredicted, likely reflecting the sparse square-planar coverage in the training set.

\begin{figure}[ht!]
    \centering
    \includegraphics[width=\linewidth]{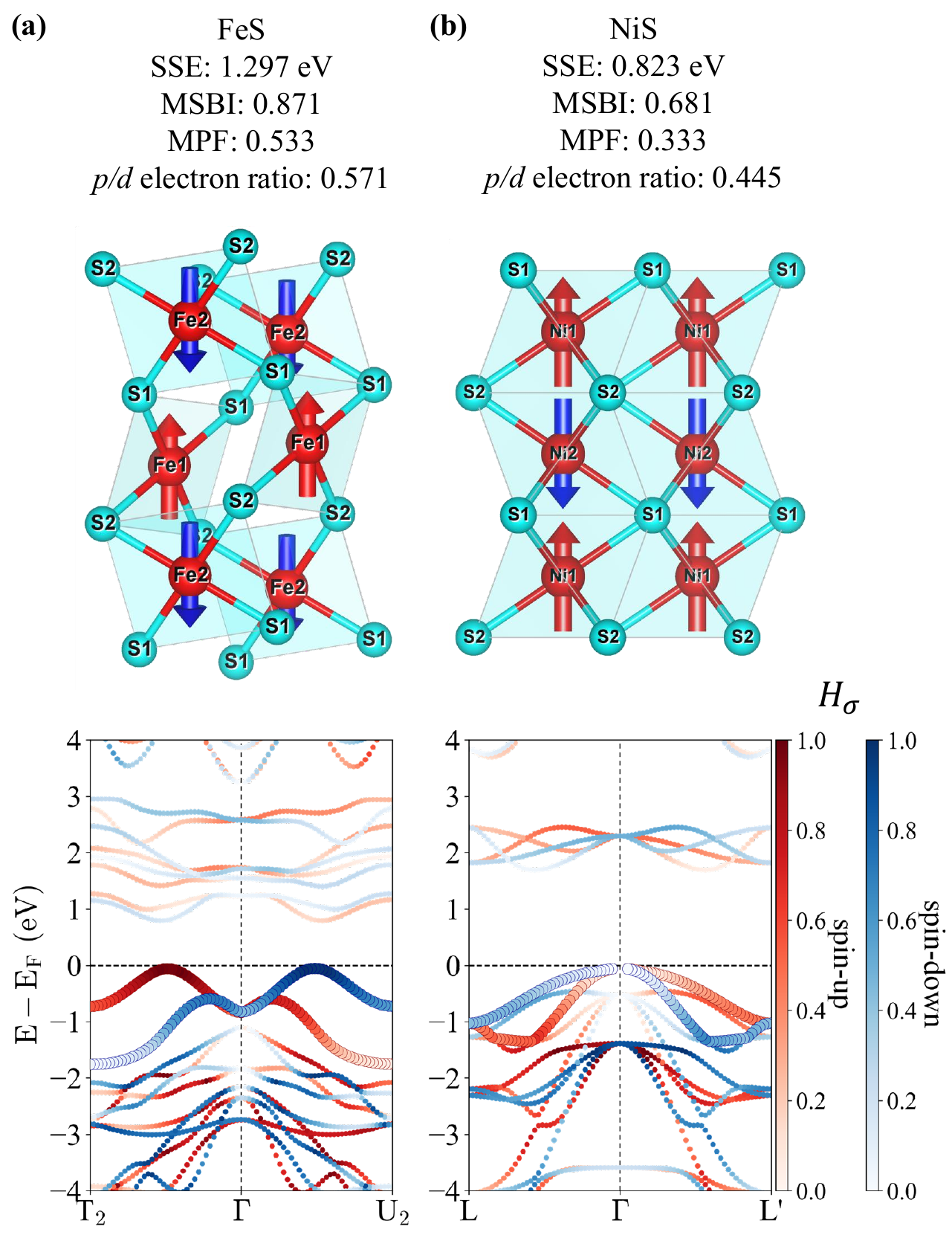}
    \caption{DFT validation of BO-selected candidates. (a) Square-planar FeS and (b) NiAs-type NiS. Top panels show the relaxed crystal structures with spin order, and bottom panels show spin-resolved $p$--$d$ hybridization fatbands along $T_2$--$\Gamma$--$U_2$ for FeS and $L$--$\Gamma$--$L'$ for NiS. The DFT-relaxed descriptors are FeS (MSBI = 0.871, MPF = 0.533, $p/d$ = 0.571) and NiS (MSBI = 0.681, MPF = 0.333, $p/d$ = 0.445), and both lie in the favorable regime identified by the SHAP analysis. Band opacity scales with $H_\sigma$, and the band pair used for SSE evaluation is highlighted.
    }
    \label{fig:validation}
\end{figure}

\textbf{Figure~5} presents the two primary validation cases. Square-planar FeS (Figure~5a) yields SSE = 1.297~eV, and the relaxed structure remains in the favorable descriptor regime. Its fatbands display strong spin-asymmetric $p$--$d$ hybridization near $E_F$, consistent with the electronic origin of the large splitting inferred from the training data. Square-planar FeS is not a known ground-state bulk polymorph, and the result should accordingly be read as a \emph{transferable motif hypothesis}: the square-planar Fe--S coordination environment itself supports giant spin splitting, wherever it can be stabilized. In the descriptor-space parameterization used here, the unit of discovery is the motif, not the bulk phase.

NiAs-type NiS (Figure~5b) provides the more immediately actionable validation case. DFT gives SSE = 0.823~eV (predicted 0.893~eV, 8.5\% error), and the relaxed bands display the same spin-asymmetric hybridization mechanism. NiAs-type NiS has been independently confirmed as a $g$-wave altermagnet from symmetry-based considerations\cite{mandal2025deterministic} and belongs to an experimentally established structural family with reported solvothermal, colloidal, and solid-state syntheses\cite{tan2019solvothermal,liu2014controlled,sun2014phase}. Epitaxial growth of $\alpha$-NiS is readily accessible: domain-matching epitaxy on $c$-plane sapphire yields a residual mismatch of only $-0.1\%$, and isostructural NiAs-type compounds---including CrSb, whose altermagnetic band splitting has been confirmed by ARPES on epitaxial films\cite{reimers2024direct}---have been grown epitaxially by MBE and sputtering for over three decades\cite{tanaka1994epitaxial}. The $\alpha$-NiS polymorph is kinetically persistent at room temperature despite being thermodynamically metastable below 652~K\cite{wang2006kinetics}, and epitaxial substrate clamping should further suppress the $\alpha \to \beta$ transition by inhibiting the accompanying 2.8\% volume expansion\cite{mcwhan1972pressure}. DFT validation results for the remaining candidates---square-planar CoO and CrS, and NiAs-type CoS and FeAs---are provided in the Supporting Information.

Taken together, the validated candidates fall into two distinct categories. NiAs-type $\alpha$-\ce{NiS} serves as a quantitative cross-validation of the framework: its altermagnetic character has been independently established through symmetry-based analysis by Mandal et al.\cite{mandal2025deterministic}, and the convergence of two methodologically distinct approaches on the same candidate is strong evidence that the learned descriptor manifold captures physically real structure--property relationships rather than statistical artifacts. The more consequential discovery output consists of three candidates with no prior altermagnetic identification: square-planar FeS (1.297~eV), octahedral CoS (1.103~eV), and octahedral FeAs (1.089~eV), all matching or exceeding the SSE of \ce{CrSb} ($\sim$1.2~eV)\cite{reimers2024direct,li2025large}. The octahedral systems adopt coordination geometries well-represented in known binary chemistry and are chemically plausible targets for solid-state or solvothermal synthesis. Square-planar FeS is most productively interpreted as a transferable motif hypothesis: the square-planar Fe--S coordination environment supports giant SSE, and epitaxial stabilization or interface-engineering strategies represent viable routes to realizing this coordination experimentally.


\section{Conclusion}
\label{sec:conclusion}

The central claim of this work is that altermagnet design does not need to inherit the binary character of the symmetry analysis that underwrites it. Magnetic point group classification tells us whether a crystal is \emph{allowed} to host altermagnetism\cite{cheong2025altermagnetism,radaelli2024tensorial}; what governs functional performance is how \emph{strongly} the allowed splitting is realized, and this quantity is continuous. We have shown that the underlying sublattice symmetry breaking itself can be promoted to a continuous, differentiable scalar---the Motif Symmetry-Breaking Index (MSBI)---that measures, in the fixed crystal frame, how far the two magnetic motifs deviate from the two spatial relations ($\mathcal{PT}$-protecting identity and inversion) whose presence would enforce spin degeneracy. Once symmetry breaking is made continuous, altermagnet design collapses from a classification problem into a well-posed optimization problem on a small number of physically interpretable axes.

Interpretable machine learning on 3,851 binary structures generated with MatterGen\cite{zeni2025generative} and labeled by spin-polarized DFT exposes that optimization problem in a three-axis form dominated by one variable and modified by two others. MSBI alone carries 23.2\% of the total SHAP\cite{lundberg2017unified} attribution of the XGBoost\cite{chen2016xgboost} surrogate---more than three times any competing descriptor---and sets a sharp geometric threshold (MSBI $> 0.5$) above which SSE systematically exceeds 0.4~eV; it is tunable through lattice distortions or symmetry-lowering substitutions. The motif packing fraction (MPF $> 0.20$) amplifies the resulting splitting by tuning the superexchange coupling scale\cite{anderson1959new,coey2010magnetism} and responds directly to applied pressure or isostructural substitution. The ligand-to-metal $p/d$ electron ratio ($p/d < 1$) selects chemistries with spin-asymmetric $p$--$d$ covalency through the Zaanen--Sawatzky--Allen mechanism\cite{zaanen1985band}. A controlled comparison between VO and CrSb, which share the same host lattice, shows that composition alone produces a sevenfold enhancement of SSE.

Bayesian optimization\cite{bergstra2011algorithms} over this three-axis space, followed by independent DFT validation, yields results in two categories. NiAs-type $\alpha$-\ce{NiS} (SSE $= 0.823$~eV, 8.5\% surrogate error) serves as a quantitative cross-validation: its altermagnetic character has been independently confirmed by symmetry-based analysis\cite{mandal2025deterministic}, and its synthesis via chemical-vapor transport\cite{gore2024growth}, solvothermal\cite{tan2019solvothermal,sun2014phase}, and thin-film routes\cite{liu2014controlled} is well-established, making spin-resolved ARPES on $\alpha$-\ce{NiS} an immediately accessible experimental test. The primary predictive output consists of three candidates with no prior altermagnetic identification: square-planar FeS (1.297~eV), octahedral CoS (1.103~eV), and octahedral FeAs (1.089~eV), all matching or exceeding the SSE of \ce{CrSb}\cite{reimers2024direct,li2025large}. Square-planar FeS is most productively read as a \emph{transferable motif hypothesis}: the square-planar Fe--S coordination environment supports giant spin splitting, and the appropriate experimental follow-up is epitaxial or interface-engineering stabilization, not the search for a bulk ground state.

The present framework is restricted to binary compounds with collinear magnetic order, a deliberate choice that ensures unambiguous identification of the magnetic ($M$) and non-magnetic ($X$) sublattices and permits a clean motif-pair comparison without the combinatorial complexity of multi-ligand coordination environments. The surrogate captures 70\% of SSE variance---sufficient for candidate ranking but not for replacing DFT. A coordination-aware spin-state prescreen would further improve the 75\% magnetic-stability retention rate observed here. Natural extensions include ternary and quaternary chemistries, more robust band matching based on irreducible representations, and generalization to noncollinear and topological altermagnetic phases, where spin--orbit coupling can open topological gaps and generate same-spin Weyl nodes at the altermagnetic band crossings\cite{vsmejkal2022beyond,vsmejkal2022emerging,hayami2019momentum}. Across all of these extensions, the three axes introduced here---a dominant continuous symmetry-breaking axis, an exchange-geometry axis, and a covalency axis---provide a transferable vocabulary for recasting the binary symmetry classification of altermagnets into a quantitative, differentiable design problem.


\begin{acknowledgments}
This research was supported by the Basic Science Research Programs through the National Research Foundation of Korea (NRF) funded by the Ministry of Education (RS-2020-NR049586, RS-2026-2547771120682075910001). Parts of the computational framework were initially developed in the M.Sc.\ thesis of K.C.\cite{Chun2025thesis}; the present work refines the descriptor definitions, introduces the controlled MSBI construction without Kabsch alignment, and adds the Bayesian optimization pipeline and its independent DFT validation.
\end{acknowledgments}

\section*{Conflict of Interest}
The authors declare no conflict of interest.

\section*{Data Availability Statement}
The trained model, descriptor-computation code, and crystal structure dataset are available at \url{https://github.com/glgkghh7539/Altermagnet-inverse-design} and archived on Zenodo (DOI: 10.5281/zenodo.19488475).

\bibliography{acs-template}

\end{document}